# Trans-pars-planar illumination enables 200° ultra-wide field pediatric fundus photography for easy examination of the retina

**DEVRIM TOSLAK[1,2], FELIX CHAU[3], MUHAMMET KAZIM EROL[2], CHANGGENG LIU[1], R. V. PAUL CHAN[3], TAEYOON SON[1,4], AND XINCHENG YAO[1,3,*]**

[1]*Department of Bioengineering, University of Illinois at Chicago, Chicago, IL, USA*
[2]*Department of Ophthalmology, Antalya Training and Research Hospital, Antalya, Turkey*
[3]*Department of Ophthalmology and Visual Sciences, University of Illinois at Chicago, Chicago, IL, USA*
[4]*Biolight Engineering LLC, Hinsdale, IL, USA.*
[*]*xcy@uic.edu*

**Abstract:** This study is to test the feasibility of using trans-pars-planar illumination for ultra-wide field pediatric fundus photography. Fundus examination of the peripheral retina is essential for clinical management of pediatric eye diseases. However, current pediatric fundus cameras with traditional trans-pupillary illumination provide a limited field of view (FOV), making it difficult to access the peripheral retina adequately for a comprehensive assessment of eye conditions. Here, we report the first demonstration of trans-pars-planar illumination in ultra-wide field pediatric fundus photography. For proof-of-concept validation, all off-the-shelf optical components were selected to construct a lab prototype pediatric camera (PedCam). By freeing the entire pupil for imaging purpose only, the trans-pars-planar illumination enables a 200º FOV in a snapshot fundus image, allowing easy visualization of both the central and peripheral retina up to the ora serrata. A low-cost, easy-to-use ultra-wide field PedCam provides a unique opportunity to foster affordable telemedicine in rural and underserved areas.

## 1. Introduction

Pediatric eye diseases such as retinopathy of prematurity (ROP) and retinoblastoma can affect both the central and peripheral retina. Therefore, wide field fundus examination is essential for screening, diagnosis and treatment evaluation of pediatric eye diseases [1, 2]. Retinopathy of prematurity (ROP) is a public health problem worldwide [3, 4]. There are more than 450,000 preterm deliveries in the USA each year [5], accounting for >10% of the ~3.9 million newborns annually [6]. In the USA alone, 400–600 infants become legally blind due to ROP each year. Globally, at least 50,000 children are blinded because of ROP each year [7]. Prompt screening and early diagnosis are essential steps to prevent visual impairment and blindness due to ROP [8]. If it could be diagnosed promptly, most of ROP caused visual losses are preventable. Laser photocoagulation and intravitreal injection of vascular endothelial growth factor (VEGF) antibodies have been approved for ROP treatment [5]. However, routine ROP screening is challenging, particularly in underserved areas and developing countries, where the access to both expensive instruments and skilled ophthalmologists is limited [9].

ROP is caused by abnormal development of retinal blood vessels in premature infants. In a healthy gestation period, retinal vascular development starts from the center of the retina, continues during pregnancy and reaches the peripheral retina after the birth. Preterm birth may disrupt the normal vascularization process, predominantly in the peripheral retina [6]. Therefore, wide field fundus examination is needed to evaluate vascular abnormality of the peripheral retina. The current gold standard for ROP screening is conventional binocular indirect ophthalmoscopy (BIO) with scleral depression [10-12], which is a time-consuming procedure that is painful for the patient and stressful for the ophthalmologist. Emerging digital pediatric fundus cameras, such as Retcam (Natus Medical Inc, Pleasanton, CA), Panocam (Visunex Medical Systems, Fremont, CA), and ICON (Phoenix Medical Systems) have improved the clinical management of pediatric patients. However, the FOV of these fundus cameras with traditional trans-pupillary illumination has been limited at 130° [1], making it difficult to access the peripheral retina adequately for a comprehensive assessment of eye conditions [13-15].

It is technically difficult to construct wide field fundus imagers, due to the complexity of illumination and imaging mechanisms. Trans-pupillary illumination allows limited FOV in a snapshot image because only the central part of the pupil can be used for collecting image light; while the periphery area of the pupil has to be used for delivering illumination light [16]. Based on the Gullstrand-Principle of fundus photography [17], the observation and illumination light beams have to be separated from each other. Otherwise, trans-pupillary illumination may cause severe light reflections from the cornea and crystalline lens, which can be multiple orders of magnitude higher than the useful signal from the retina. Sophisticated optical design and delicate system construction increase the instrument complexity and cost of the pediatric fundus camera. With the limited FOV, it is time-consuming for clinical examination of the retina, and technically difficult to access zone III for a comprehensive ROP management [18]. Scanning laser ophthalmoscope (SLO), such as Optos (Optos, Dunfermline, UK) has been demonstrated to provide wide field fundus photography. By combing two or more laser wavelengths, color fundus SLO photography is practical. However, sophisticated scanning device has to be involved, compared to traditional snapshot fundus cameras. Although a 'flying baby position' has been demonstrated to use Optos for pediatric imaging [19], its clinical deployment is still difficult.

The pars plana is a smooth, posterior part of the ciliary body. The pars plana lacks muscle, blood vessels and pigmentation [20]. Therefore, it is more transparent than other scleral areas, making it an alternative location for delivering light into the eye [21-23]. Trans-scleral illumination has been successfully demonstrated for retinal imaging of adult eyes [24, 25]. However, the previous effort for exploring trans-scleral illumination in pediatric fundus imaging was failed [26]. In adults, the pars plana is a ~ 4 mm posterior part of the ciliary body [27, 28]. However, the dimension of the pars plana is closely correlated with postconceptional age in pediatric patients. The mean pars plana width in full-term infants is between 1.5-2 mm and will develop to ~4 mm after 1-2 years [29, 30]. Therefore, careful control of the illumination spot size and accurate identification of the pars plana location are essential factors to optimize trans-pars-planar illumination in pediatric fundus photography. In this study, we demonstrate the first successful implementation of contact-mode trans-pars-planar illumination in pediatric fundus photography, with 200° FOV in a single-shot image, allowing easy examination of the peripheral retina up to the ora serrata, i.e., the far end of the retina.

## 2. Materials and methods

### 2.1 Experimental setup

Figure 1A illustrates the optical layout of a lab prototype pediatric camera (PedCam). All off-the-shelf optical components were selected to construct the handheld PedCam for proof-of-concept validation (Fig. 1B). The trans-pars-planar illuminator in Fig. 1A consists of an optic fiber with 1.8 mm diameter and 0.65 numerical aperture (NA) (FO-DIAG-PROBE-1.8, Salvin Dent. Spec. Inc., Charlotte, NC) and a 565 nm light emitting diode (LED) (M565L3, Thorlabs Inc, Newton, NJ). The first component of the imaging system is a contact ophthalmoscopy lens (OL) (HRX vit, Volk Optical Inc, Mentor, OH). The contact OL consists of two elements, a meniscus lens and a condensing lens. The removable design of the contact OL in the prototype PedCam enables easily sterilization of the meniscus lens contacting to the eye. The contact OL produces an aerial retinal image in front of the relay optics (Fig. 1D). The relay optics consists of three lenses, i.e., a bi-convex lens I with focal length f = 60.0 mm (LB1596, Thorlabs Inc., Newton, NJ), a planoconcave lens II with f = -50.0 mm (LC1715, Thorlabs Inc., Newton, NJ) and another bi-convex lens III with f = 30.0 mm (EO-89187, Edmund Optics Inc., Barrington, NJ). Zemax simulation (OpticStudio, Zemax LLC, Kirkland, WA) was conducted to select these lenses to minimize the system dimension and optical aberration (Fig. 1E). In coordination with the relay optics and a camera lens with f=12 mm (EO-33303, Edmund Optics Inc., Barrington, NJ), the aerial retinal image is relayed to the camera sensor (Fig. 1E). The digital camera (EO-6412 LE, Edmund Optics Inc., Barrington, NJ) has a frame resolution 3088 x 2076 pixels with 2.4 x 2.4 μm pixel size. For capturing images presented in the article, the aperture of the camera lens was set to F1.8.

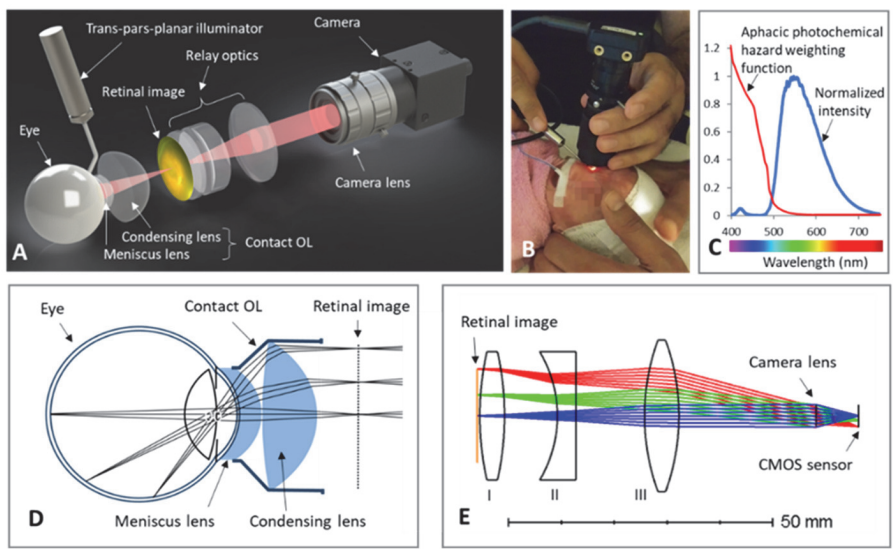

Fig. 1. (A) Optical layout of the prototype PedCam. All off-the-shelf optical components were used for this prototype. (B) Photographic illustration of the prototype PedCam for fundus imaging. (C) Light spectrum, normalized intensity of the light source and ISO 15004-2: 2007 aphacic photochemical hazard weighting function. (D) The contact OL, consisting of a meniscus lens contacted to the cornea and a condensing lens, is used to produce an aerial retinal image at ~3 mm from the last surface of the lens set. (E) The relay optics in Fig. 3A, consisting of a bi-convex lens I, a planoconcave lens II and another bi-convex lens III, working with the camera lens together to relay the aerial retinal image to the camera sensor.

## 2.2 Human subjects

This study was approved by the Institutional Review Board of the University of Illinois at Chicago and was in compliance with the ethical standards stated in the Declaration of Helsinki. Two patients with ROP and one patient with previously diagnosed retinoblastoma were used for functional validation of the PedCam with trans-pars-planar illumination. As shown in Fig. 1B, the subject head was held by an ophthalmic technician to keep him/her still during the image recording. Topical tetracaine 0.05% was used for anesthesia while cyclomydril, phenylephrine, and/or tropicamide were used for pupil dilation. General anesthesia was also used for the retinoblastoma patient as part of a routine exam under anesthesia. An eye speculum was used to keep the eyelids open during the recording. GenTeal Lubricant Eye Gel (Alcon Laboratories, Fort Worth, TX) was applied between the contact OL and the eye. All images were captured by pediatric ophthalmologists using the PedCam with trans-pars-planar illumination. Considering the ~3 mm ciliary body length in newborns [31, 32], the illumination probe was placed ~1.5 mm away from the corneoscleral limbus to cover the pars plana. During image focusing adjustment, live video images were streamed to the computer for continuous monitoring. Image focusing was performed manually by rotating the focusing ring on the camera lens. Slightly moving the illuminator back and forth from the limbus, the optimal illumination location, i.e. the pars plana, could be identified based on the image quality [23]. For reaching the ora serrata region, the imaging probe was slightly tilted from the axis.

*2.3 Light safety*

ISO 15004-2 is the standard guidance for assessing light safety of ophthalmic devices. Although there is no special requirement for pediatric patients, conservative estimation was carefully used to ensure the safety of newborns. For light safety, both photochemical and thermal hazards of the retina were quantitatively evaluated. It is known that the photochemical hazard to biological tissues is primarily related to the blue light absorption (Fig. 1C). Therefore, a 565 nm LED was selected as the light source to minimize the photochemical hazard risk to the retina. The previous study has reported that the thickness of the sclera is ~ 0.5 mm [33]. The transmission of the sclera in visible wavelength is 10% - 30% [34]. 30% transmission was used for conservative assessment of the light safety. According to the ISO 15004-2 standard [35], a maximum of 10 $J/cm^2$ weighted irradiance is allowed on the retina without photochemical hazard concern. The weighted irradiance in Fig. 1C was calculated using the photochemical hazard weighting function provided in the ISO 15004-2 standard [35]. For the 7 mW maximum light power, the weighted power was 0.117 mW based on the spectral measurement with a spectroradiometer (PR-670, Photo Research, Chatsworth, CA). For conservative estimation of the worst case, assuming all light directly reaches to the retina behind the illuminated sclera area, the illuminated retinal area was estimated as 5.48 $mm^2$, the maximum allowed exposure time is [35]:

$$t = \frac{10 \text{ J/cm}^2}{0.117 \text{ mW} \times 30\%/5.48 \text{mm}^2} = 4.3 \text{ hours} \quad (1)$$

The maximum weighted power intensity allowed on the sclera without thermal hazard concern is 700 $mW/cm^2$ [35]. The weighted power intensity of the illumination light in this study was 127 $mW/cm^2$, which was more than five times lower than the maximum limit. Therefore, there was no thermal hazard concern.

## 3. Results

Figure 2A and Fig. 2D shows representative fundus images of ROPs captured with the 200° prototype PedCam under topical anesthesia. Figure 2B illustrates a schematic diagram of the eye, showing the locations of the pars plana, ora serrata, peripheral retina and equator of the globe. From the ora serrata to ora serrata, the whole retina covers 230° internal eye angles [36]. The peripheral retina refers to a ring shape region between the ora serrata and the equator, accounting for a ~50° region of the 230° retina (Fig. 2B). Figure 2C is a schematic diagram of the left ocular fundus, illustrating zones I, II, III for ROP classification. As shown in Fig. 2A, the single snapshot image can cover the retina from the optic disc to the ora serrata. By delivering the trans-pars-planar illumination light from the temporal (Fig. 2D1) and nasal (Fig. 2D2) sides, respectively, fundus examination from the ora serrata to ora serrata can be readily implemented.

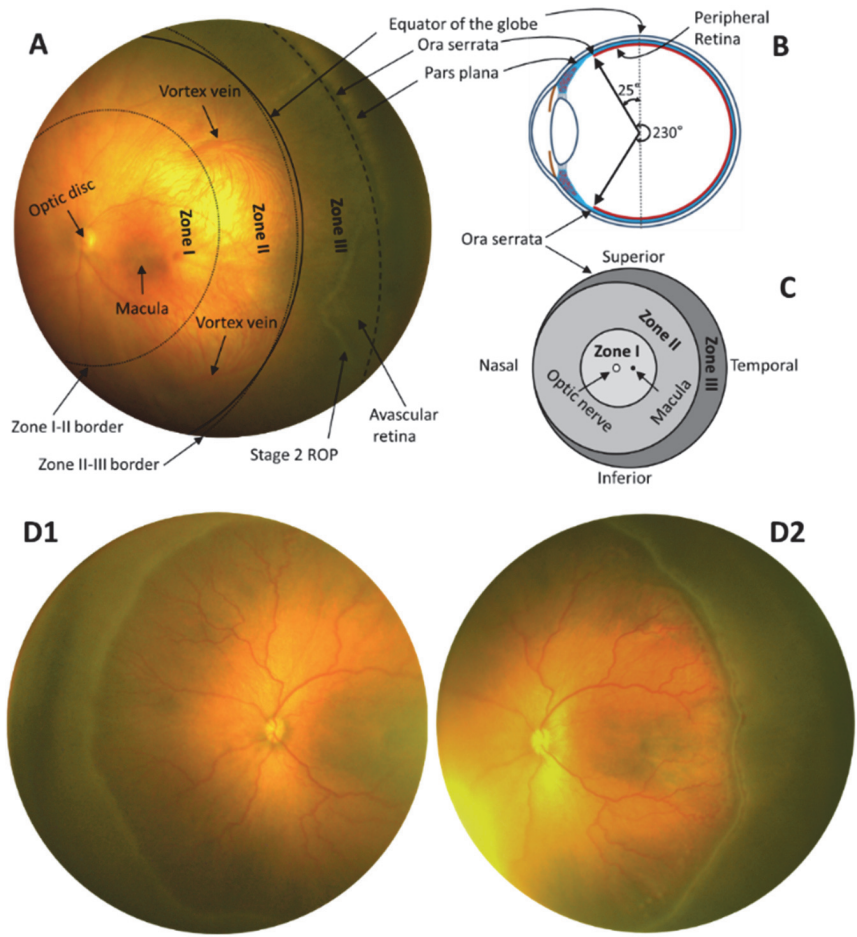

Fig. 2. (A) Representative fundus images captured with the prototype PedCam from the left eye of a patient with zone III stage 2 ROP. (B) Schematic diagram of the eye showing the location of the pars plana, ora serrata, peripheral retina, and equator of the globe. (C) Schematic diagram of the left ocular fundus, illustrating retinal zones for ROP classification. (D) Representative fundus images captured with the prototype PedCam from the left eye of a patient with plus disease, zone II stage 2 ROP in the nasal region and zone II stage 3 ROP in the temporal region, with trans-pars-planar illumination light delivered from the temporal (D1) and nasal (D2) sides, respectively.

In addition to the ROP imaging in Fig.2, we also conducted a comparative study of a patient previously diagnosed with retinoblastoma using the prototype PedCam and a clinical RetCam (Fig. 3). Retinoblastoma is the most common primary malignant intraocular tumor in children and can be life threatening without prompt medical intervention [37]. Figure 3 shows representative images of the patient under topical and general anesthesia. A treated retinoblastoma lesion surrounded by photocoagulation scars was observed in Fig. 3A. Figure 3A shows the fundus image captured with the prototype PedCam. By slightly tilting the imaging system relative to the visual axis of the eye, the ora serrata and pars plana region were unambiguously observed in the superior temporal quadrant (Fig. 3B). For FOV comparison, two images captured with the prototype PedCam (200°) and Retcam (130°) from the same subject are overlapped in Fig. 3D.

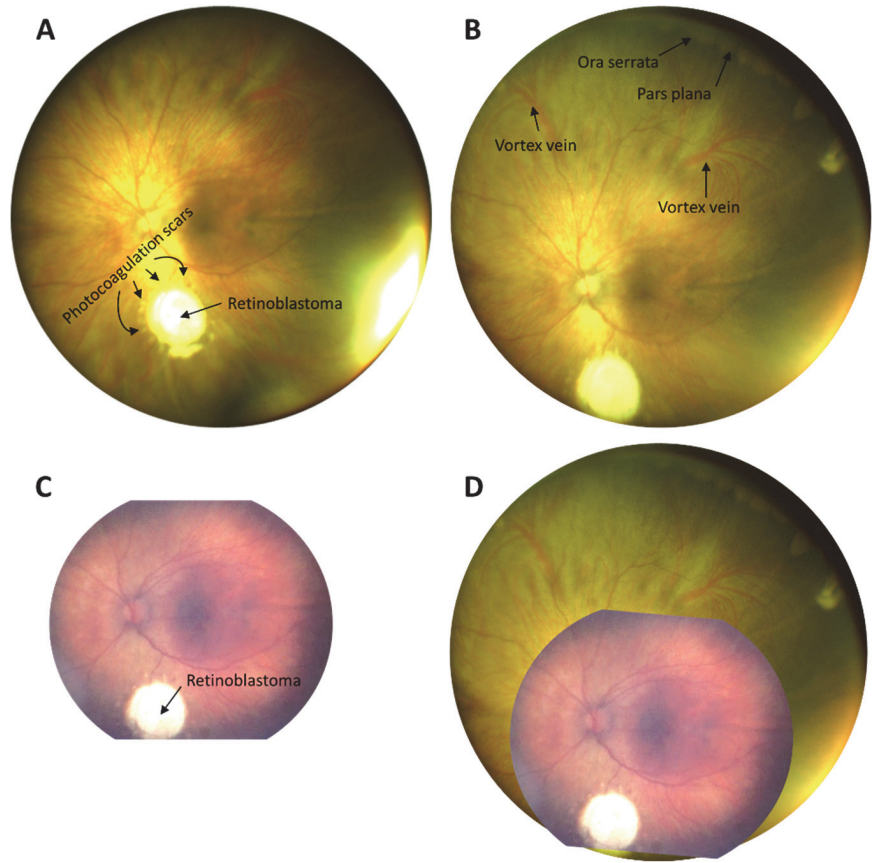

Fig. 3. Fundus images of a patient previously diagnosed with retinoblastoma. (A) Central view of the fundus captured with the prototype PedCam2 in Fig. 1. A treated retinoblastoma lesion surrounded by photocoagulation scars was observed below the optic disc. (B) Vortex vessels, ora serrata and pars plana were observed by slightly tilting the axis of the imaging system for peripheral imaging. (C) Fundus image of the same eye captured by clinical Retcam. (D) Overlapping illustration of two images captured by the 200º prototype PedCam and 130º clinical Retcam.

## 4. Discussion

In summary, we demonstrated the feasibility of using trans-pars-planar illumination for ultra-wide field fundus photography. To the best of our knowledge, this is the first successful demonstration of using contact-mode trans-pars-planar illumination in pediatric fundus photography. Contact-free trans-pars-planar illumination has been demonstrated for wide field fundus photography [23]. Contact-free imaging requires careful cooperation of the patients for retinal alignment and focusing, because poor fixation and eye movement may significantly affect the image quality [38]. Therefore, the contact-free modality is not practical for uncooperative pediatric patients, particularly newborns. Contact-mode imaging is required for pediatric patients to minimize the effect of eye movements on retinal imaging [39].

All off-the-shelf optical components were selected to construct the lab prototype PedCam (Fig. 1) for the proof-of-concept demonstration. Both patients with ROP and retinoblastoma were used for functional validation of the prototype PedCam. Compared to the limited FOV of BIO with scleral depression, the trans-pars-planar illumination enabled a 200° FOV in single snapshot fundus image without the need of scleral depression, allowing easy visualization of peripheral retina up to the ora serrata. Because the trans-pars-planar illumination frees the whole pupil area for imaging purpose only, the complexity of the imaging system is significantly simplified compared to the sophisticated illumination and imaging optics in traditional fundus cameras with trans-pupillary illumination [16]. The total material cost of the prototype PedCam is ~$2,500. Compared to the expensive cost of current pediatric fundus cameras, the low-cost, ultra-wide field PedCam may provide an affordable solution to advance telemedicine in rural and underserved areas.

The 130° RetCam is the most common device used for digital photography of ROP. Without scleral depression, the maximum field of regard, i.e., total FOV available with the RetCam is limited to the equator of the eye globe [40]. In other words, current digital photography cannot provide sufficient FOV for a comprehensive examination of ROP and other pediatric eye diseases [41]. Trans-pars-planar illumination enables 200° FOV in a single snapshot image, which is wider than the total FOV covered by 5 mosaic images (central, inferior, superior, temporal and nasal angles) provided by the 130° RetCam for telemedicine examination of ROP [41]. To capture three 200° images including central, temporal and nasal angles, the whole imaging session required less than 5 minutes including the procedures for illumination alignment and image focusing adjustment. With the capability of imaging peripheral retina up to the ora serrata, the PedCam provides a unique opportunity for comprehensive eye examination of pediatric patients, allowing digital fundus photography for routine clinical deployment and affordable telemedicine of ROP and other pediatric eye diseases.

Our current prototype PedCam with all-off-shelf optical parts has some limitations. For the current prototype, a 3-4 mm pupil size is typically required for visualizing the whole retina from the center to the ora serrata. In addition, effective pupil size for off-axis observation of peripheral region is reduced to affect light efficiency, i.e. vignetting effect as shown in Fig. 2 and Fig. 3. We are currently pursuing custom-designed optics to lower the pupil size required for imaging the peripheral retina without the need of pharmacological pupil dilation. We also anticipate to optimize optical design and to employ digital compensation to overcome the vignetting problem in ultra-wide field fundus imaging.

## 5. Conclusion

In conclusion, a trans-pars-planar illumination based PedCam has been demonstrated to enable a 200º FOV for ultra-wide field pediatric fundus photography. An ultra-wide field PedCam provides a unique opportunity to foster clinical deployments of digital fundus photography for pediatric eye disease management. T low-cost, easy-to-use, portable PedCam also provides an affordable solution to advance telemedicine in rural and underserved areas.

## Funding

This research was supported in part by NIH grants R43 EY028786, R01EY029673, R01 EY030101, P30 EY001792; by unrestricted grant from Research to Prevent Blindness; by Richard and Loan Hill endowment.

## Acknowledgment

The authors acknowledge Mr. David Le for a proofreading of this manuscript.

## Disclosure

D. Toslak and X. Yao have patent applications.